\documentclass[journal]{IEEEtran}
\usepackage{epsfig}
\usepackage{graphicx}
\usepackage{amsmath}
\usepackage{amssymb}
\usepackage[linesnumbered,boxed]{algorithm2e}
\usepackage{color}
\usepackage{multirow}
\usepackage{rotating}
\newcommand{\etal}{\textit{et al}.}

\newcommand{\eg}{\textit{e}.\textit{g}. }

\begin{document}

\title{A learning-based view extrapolation method for axial super-resolution}

\author{\IEEEauthorblockN{Zhaolin Xiao\IEEEauthorrefmark{1}, Jinglei Shi\IEEEauthorrefmark{2}, Xiaoran Jiang\IEEEauthorrefmark{2}, Christine Guillemot\IEEEauthorrefmark{2} (IEEE Fellow)}
\thanks{\IEEEauthorrefmark{1} Xi'an University of Technology, Xi'an 710048, China.}
\thanks{\IEEEauthorrefmark{2} Institut National de Recherche en Informatique et en Automatique (INRIA), Rennes 35000, France.}
\thanks{}
}

\markboth{Submitted to Journal}%
{Shell \MakeLowercase{\textit{et al.}}: Bare Demo of IEEEtran.cls for IEEE Journals}
\maketitle

\begin{abstract}
Axial light field resolution refers to the ability to distinguish features at different depths by refocusing. The axial refocusing precision corresponds to the minimum distance in the axial direction between two distinguishable refocusing planes. High refocusing precision can be essential for some light field applications like microscopy. In this paper, we propose a learning-based method to extrapolate novel views from axial volumes of sheared epipolar plane images (EPIs). As extended numerical aperture (NA) in classical imaging, the extrapolated light field gives re-focused images with a shallower depth of field (DOF), leading to more accurate refocusing results. Most importantly, the proposed approach does not need accurate depth estimation. Experimental results with both synthetic and real light fields show that the method not only works well for light fields with small baselines as those captured by plenoptic cameras (especially for the plenoptic 1.0 cameras), but also applies to light fields with larger baselines.
\end{abstract}


\IEEEpeerreviewmaketitle

\section{Introduction}

\IEEEPARstart{L}{ight} field imaging has become popular in the last years, due to its potential for a variety of applications. Light field imaging enables post-capture digital refocusing which is an interesting functionality for example in computational photography and microscopy. Digital refocusing can be easily implemented by shifting and adding the sub-aperture images of the light field \cite{Isaksen2000}. Reviews on light field imaging, cameras and applications can be found in \cite{Wu17overview,Ihrke16overview,Wetzstein11overview,Levoy06overview}.
\begin{figure}[t]
\centering
\includegraphics[width=\linewidth]{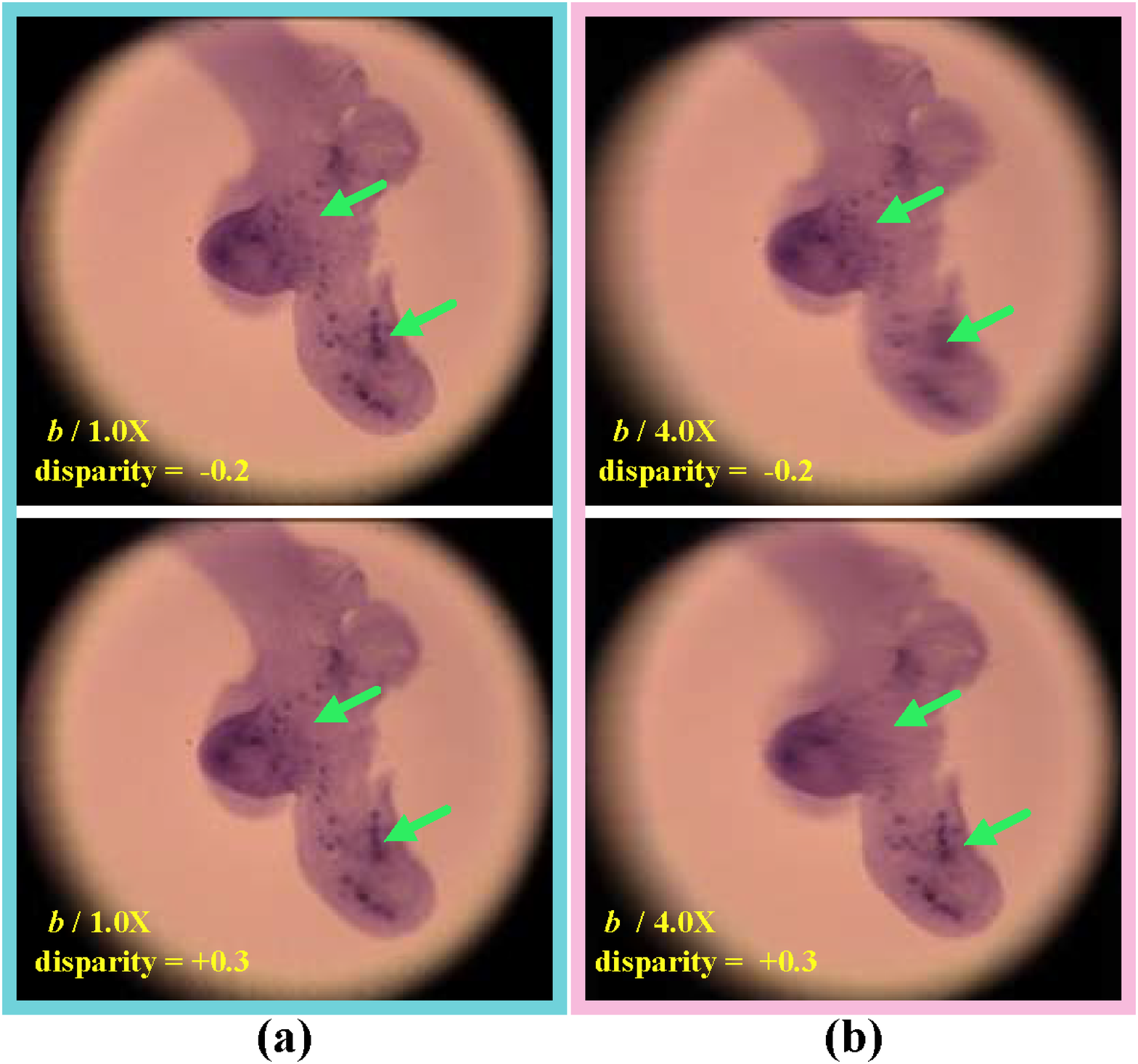}
\caption{Light field refocusing precision enhancement. (a) Original refocusing at two axial positions corresponding to a given disparity, hence} to a specific focal plane when applying digital refocusing. The parameter $b$ represents the baseline in the angular plane. (b) Enhanced refocusing with the same focus setting as in (a), from which we can see clear differences with a
$4.0X$ computationally enlarged baseline.
\label{fig:1}
\end{figure}
Many light field acquisition devices have been designed in the past two decades, exhibiting a trade-off between the spatial and the angular resolution of the captured data. For example, plenoptic cameras, plenoptic 1.0 \cite{Ng2005LightFP} or plenoptic 2.0 cameras \cite{RaytrixWeb}, due to optical and sensor limitations, sacrifice spatial resolution for increased angular resolution, leading to a lower spatial resolution compared to traditional 2D cameras. In traditional 2D image or video processing, the low resolution problem can be alleviated by applying spatial or temporal super-resolution, \eg exploiting intra-frame correlations \cite{YiTPAMI2020} and inter-frame correlations \cite{YiTCSVT2020}. But, increasing the angular light field resolution is also an important issue, as a limited resolution may limit the practical use of light fields in real applications. For this reason, various methods have been developed to achieve a better spatial and angular resolution trade-off ~\cite{Bishop2012,Wanner-SR2013,Guillemot2020}.

In this paper, we focus instead on enhancing the light field axial resolution by improving the refocusing precision. In real systems, due to limited spatial resolution and baseline, the number of distinguishable focal planes is limited in the axial direction. Please notice that, while the light field baseline often denotes the spacing between two adjacent views (or cameras), here the term {\em baseline} will refer to the spacing between two farthest views (or cameras).

In this paper, we present a learning-based axial refocusing precision enhancement framework by computationally extending the virtual light field baseline.
Most importantly, the proposed solution does not require an explicit and accurate depth estimation.
The refocusing precision can be essential for some applications, \eg light field microscopy~\cite{Prevedel2014Simultaneous,Palmieri2019}, and light field particle image velocimetry (LF-PIV)~\cite{Belden_2010,Skupsch13-PIV}.
Figure~\ref{fig:1}-(a) shows that, when using the shift-and-add refocusing method on the original light field, one can hardly see refocusing differences. On the contrary, Figure~\ref{fig:1}-(b) shows that, by enhancing the refocusing precision, we can better distinguish the objects at different axial positions.
Please notice that we used the same focal plane setting for both Figures~\ref{fig:1}(a) and (b).
We propose a learning-based solution operating on axial volumes of EPIs to extrapolate structured light field views, which we called EPI shearing and extrapolation network (EPI-SENet). We introduce a forward and backward shearing strategy on 3D EPI volumes to avoid explicit depth estimation.
Experimental results show that the method not only works well for light fields with small baselines as those captured by plenoptic cameras (especially for the plenoptic 1.0 cameras), but also applies to light fields with larger baselines. We show that the proposed solution can effectively extend the baseline to $4\times$ larger, and that extended baseline gives re-focused images with a shallower depth of field leading to more precise refocusing.

\section{Related work}

In classical optical design, increasing the numerical aperture (NA) decreases the DoF, but leads to a higher axial resolution (also called depth resolution)~\cite{Murphy2012Fundamentals}. Many solutions have been proposed to deal with the narrow DoF problem, such as image deblurring based on 3D PSF modeling and all-in-focus image fusion from multiple axial scans~\cite{Mart2009Chapter}. In the object space, Chen \etal ~\cite{Jin2017Distance} and Hahne \etal ~\cite{Hahne2016Refocusing} proposed different optical models to accurately measure the distance of the object plane based on a geometric analysis of plenoptic 1.0 and standard plenoptic cameras respectively. Furthermore, Hahne \etal ~\cite{Hahne2016Refocusing} derived the distance to the refocused object plane and its corresponding DoF for different light fields, which has been experimentally verified by placing objects at the predicted distances. Instead of exploring the DoF, we focus on enhancing the axial resolution by computationally extending the light field baseline. In light field imaging, the axial resolution can be enhanced by increasing spatial resolution or by virtually extending the baseline. Extending the light field baseline gives re-focused images with shallower DoF (as extending the numerical aperture in classical imaging), leading to more accurate re-focusing. Below we review light field super-resolution and extrapolation methods that could contribute to axial resolution enhancement.

\subsection{Enhancing light field resolution}
In ~\cite{Bishop2009}, Bishop and Favaro model the image formation process of lenslet-based light field cameras, and model both the main lens and the micro-lens blur formation. The model gives the relationship between spatial resolution and defocus blur. The authors in \cite{Bishop2009} propose a method to estimate depth dependent point spread functions (PSF), which are then used to solve the spatial super-resolution in a Bayesian inference framework. Broxton \etal ~\cite{Broxton2013} propose a 3D deconvolution method to produce higher spatial resolution for light field microscopy. Furthermore, in order to make the effective spatial resolution more uniform along the axial direction, Cohen \etal  ~\cite{Cohen2014} suggest precisely controlling the shape of the light field PSF by placing two phase masks in the back focal plane of the objective lens and in the micro-lenses apertures respectively.

A patch-based technique is proposed in \cite{Mitra2012} where high-resolution 4D patches are estimated using a linear minimum mean square error (LMMSE) estimator assuming a disparity-dependent Gaussian Mixture Model (GMM) for the patch structure. Given the estimated depth maps, the authors in \cite{Wanner-SR2013} first estimate depth by analyzing the 1D structures in EPIs and use a variational optimization framework to spatially super-resolve the light field and then increase its angular resolution. A CNN is used in ~\cite{Guillemot2020} to learn a model of correspondences between low- and high-resolution data in subspaces of reduced dimensions.

\subsection{Extending the light field baseline}
Virtually extending the angular baseline, by view extrapolation, is another effective solution for increasing the refocusing precision. Different methods have been proposed either for view interpolation and extrapolation.
The view interpolation problem is solved in~\cite{Vagharshakyan2018} using a sparsity prior in an adapted discrete shearlet transform domain. The authors in \cite{Levin2010} first compute the focal stack from the input light field and interpolate and extrapolate novel views by de-convolution of focal stack images. The method in \cite{Shi2014} based
on a Sparse Fast Fourier Transform (SFFT) exploits sparsity in the angular dimensions of the 4D Fourier domain to recover the light field from a subset of views.
Le Pendu \etal ~\cite{LePendu2019} suggest decomposing light fields into multiple depth layers in the Fourier domain. This representation, called Fourier disparity layers (FDL), enables not only a flexible control of focusing depth, view-point, and aperture shape, but also view interpolation and extrapolation for enlarging the virtual aperture size.
Zhou \etal ~\cite{Zhou2018} suggest encoding the scene content and visibility as a set of $RGB$ multi-plane images (MPI), then they use the MPI representation for view interpolation~\cite{Mildenhall2019} and extrapolation~\cite{Srinivasan2019} from a limited set of views.

Deep neural networks have been proposed both for view interpolation and extrapolation. The authors in \cite{Kalantari2016} propose an architecture based on two CNN. The first CNN being used to estimate disparity maps between a target viewpoint and each input (corner) view of the light field. The disparity maps are then used to warp the four input views into the target view positions, and the second CNN computes the color of the target views based on the warped corner views. The authors in~\cite{Wu2017} learn a CNN to predict confidence scores of the different shears, and these scores are then used to merge the EPIs for view synthesis. Wang \etal ~\cite{Wang2018} introduce a Pseudo 4DCNN to generate dense light field from angular sparse input, instead of applying on 2D EPI image, the 4DCNN is trained to interpolate views on 3D EPI volumes. Yeung \etal ~\cite{Henry2018Fast} propose an end-to-end dense light field reconstruction framework, which uses a coarse-to-fine strategy to synthesize novel views by applying guided residual learning. Wu \etal ~\cite{Wu2019} propose a learning-based framework to angularly interpolate light fields from a sparse set of views, in which a 2D encoder-decoder network is introduced to reconstruct high resolution line-features by using sheared 2D EPIs. Relying on accurate disparity estimation, Shi \etal ~\cite{Shi2020Learning} suggest fusing a pixel-based and a feature-based view reconstruction using a learned soft mask.

The proposed view extrapolation method shares similarities with the method in ~\cite{Wu2019}, with however two major differences. First, our learning-based solution is designed to extrapolate views beyond the angular boundaries, rather than reconstructing views within the boundary. Second, our input is a 4D sequence of multiple sheared 3D EPI volumes. Instead of using a prior upsampling as in~\cite{Wu2019}, the novel angular views are directly predicted by the proposed network architecture. In addition, instead of using a pyramidal decomposition as in~\cite{Wu2019}, we use a fusion network with a learned confidence to merge the extrapolation results obtained with the multiple shearings.

\section{Axial refocusing precision and depth of field}\label{sec:3}

\subsection{Refocusing precision: definition}

\begin{figure}[h]
\centering
\includegraphics[width=\linewidth]{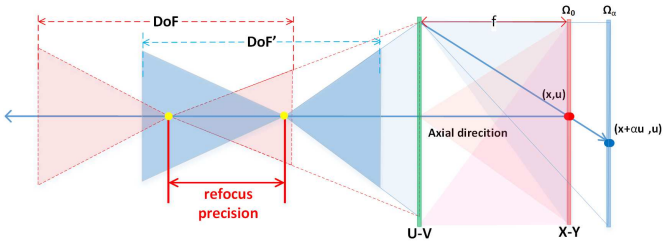}
\caption{Difference between light field DoF and refocusing precision. The pink and blue shadowed regions correspond to the definition of traditional DoF. The refocusing precision is defined as the minimum distance $Arp$ between two distinguishable adjacent focal planes in the object space.}
\label{fig:2}
\end{figure}

Let $\Omega_{0}$ be a given focal plane, $\Omega_{0}^{\alpha^{+}}$ and $\Omega_{0}^{\alpha^{-}}$ are the farthest and the nearest distinguishable planes around $\Omega_{0}$ respectively.
The refocusing precision is defined as the minimum spacing distance between two distinguishable adjacent focal planes in the object space, and can be expressed as
\begin{equation}\label{equ:equ2}
\begin{array}{l}%
\emph{Arp}\left(LF_0,\Omega_{0}\right)= \left[d\left(\Omega_{0}^{\alpha^{-}}\right), d\left(\Omega_{0}^{\alpha^{+}}\right)\right]\\
{\rm s.t.} \hspace{0.1cm} \left\|L F_{0}^{\alpha}-L F_{0}\right\|<\varepsilon, \text { if } \alpha \in\left[\alpha^{-}, \alpha^{+}\right]
\end{array}%
\end{equation}
where $d(\Omega_{0})$ stands for the axial position of the focal plane $\Omega_{0}$ in the object space, i.e. the distance between the conjugate focal plane of $\Omega_{0}$ on the object side and the camera plane $UV$ (the axial zero position). The parameter $\varepsilon$ denotes a negligible difference between the original light field $LF_0$ and the resampled one $L F_{0}^{\alpha}$.
Figure~\ref{fig:2} shows the difference between the axial refocusing precision and the DoF in the object space. It shows that the two adjacent distinguishable focus planes located at distances $d\left(\Omega_{0}^{\alpha^{-}}\right)$ and $d\left(\Omega_{0}^{\alpha^{+}}\right)$ from the camera plane do not correspond to the borders of the DoF.
Due to finite and discrete sampling of the 4D light field, the $Arp\left(LF_0,\Omega_{0}\right))$ cannot be infinitely small with a fixed spatio-angular resolution.

\subsection{Depth of Field (DoF)}

The axial re-focusing precision has some relationship with the concept of DoF.
The DOF is indeed the distance (red and blue dash lines in Fig.\ref{fig:2}), within which the object is in-focus, i.e. without optical blur. The DOF has been exaggerated compared to the dimension of the imaging system for better observation.

In the case of real light field imaging systems, and assuming the number of angular views in both dimensions is the same, i.e. $N_u=N_v$, where $N_u$ and $N_v$ are the numbers of views in the horizontal and vertical dimensions respectively, the DoF of refocused image is given by~\cite{Levoy2006LFM}
\begin{equation}\label{equ:equ4}
  DoF \approx \frac{\lambda n}{\emph{NA}^{2}} +
  \frac{N_{u} \lambda n}{2{\emph{NA}}^{2}}
\end{equation}
where $\lambda$ stands for the light wave length, $n$ represents the refraction index of medium, and $\emph{NA}$ is the numerical aperture of the entire imaging system. Theoretically, the total DoF is determined primarily by wave optics (first term of Equation (\ref{equ:equ4})) which dominates if the pixels are small enough to not limit resolution, and which corresponds to the theoretical wave DOF limit. But, for lower numerical apertures, the DoF is dominated by the geometrical optical circle of confusion (CoC) represented by the second term of Equation (\ref{equ:equ4}). The geometrical optical CoC, which is $N_{u}/2$ times larger than the wave optics term, dominates. As a consequence, the DoF of light field imaging is significantly larger than the DoF of traditional imaging with the same resolution sensor.

Please note that, in the case where the DoF of all the focal slices is non-overlapping in a focal stack, and that we are interested in the in-focus regions only, then the refocusing precision $\emph{Arp}$ is equivalent to the $DoF$. However, this assumption usually does not hold in real light field imaging systems for the two following reasons.
First, since the pixel size may not be small enough to satisfy the theoretical wave DoF limit, the actual DoF of the different slices of the focal stack may be larger and overlapping in most practical light field imaging systems, as shown in Figure \ref{fig:2}.
Second, the refocused pixel size is $N_u \times N_v$ times larger than the sensor pixel size or the diffraction-limited spot size. Here, $N_u \times N_v$ is the angular resolution.
While the DOF only measures the range of in-focus area, the axial refocusing precision measures the distance between adjacent distinguishable focus planes.

\section{Proposed view extrapolation method}

In this section, we propose a learning-based solution to computationally extend the angular baseline by extrapolating novel views. An overview of the proposed angular extrapolation framework is shown in Figure ~\ref{fig:6}. We use a forward and backward shearing strategy applied on EPI volumes, with multiple sheared candidates to better deal with possible discontinuities in EPI 1D structures. Unlike the method in ~\cite{Wu2019}, our network is designed for 4D EPI volume input, from which the network is trained to learn EPI line features and to extrapolate novel angular samples. Note that extrapolation is a more difficult task than interpolation in particular due to occlusions.

\begin{figure*}[tb]
\centering
\includegraphics[width=\linewidth]{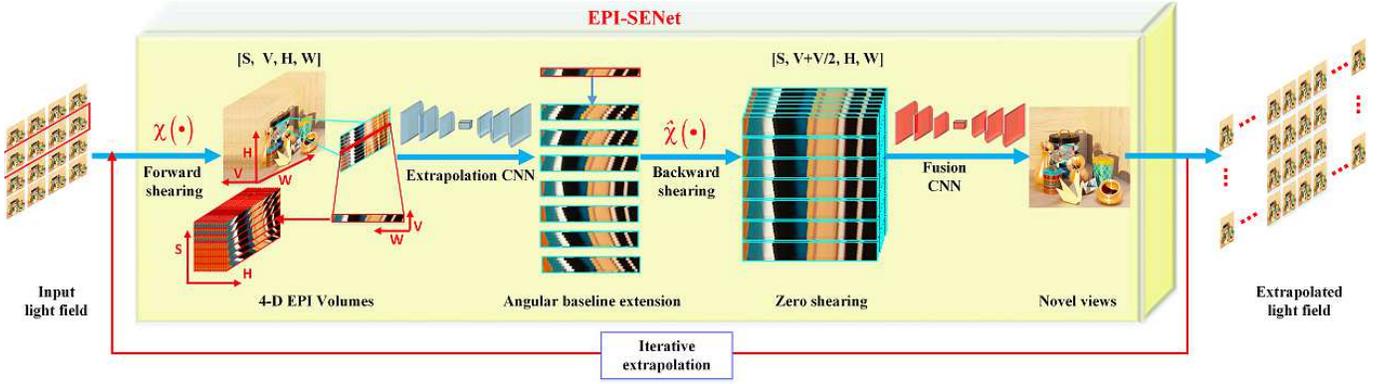}
\caption{Overview of the EPI-SENet pipeline.
The first step consists in first constructing 3D EPI volumes of dimension $[H,W,V]$ of different shears. Then the 3D EPI volumes of different shears lead to 4D EPI volumes of dimension $[H,W,V,S]$, where S denotes the number of shears. The 4D EPI volumes are then fed into the learned extrapolation network. A backward shearing is applied on the extrapolated EPI volumes, which are then fed to a second network which merges the different extrapolated candidates corresponding to the different shears.}
\label{fig:3}
\end{figure*}

\begin{figure}[tbh]
\centering
\includegraphics[width=\linewidth]{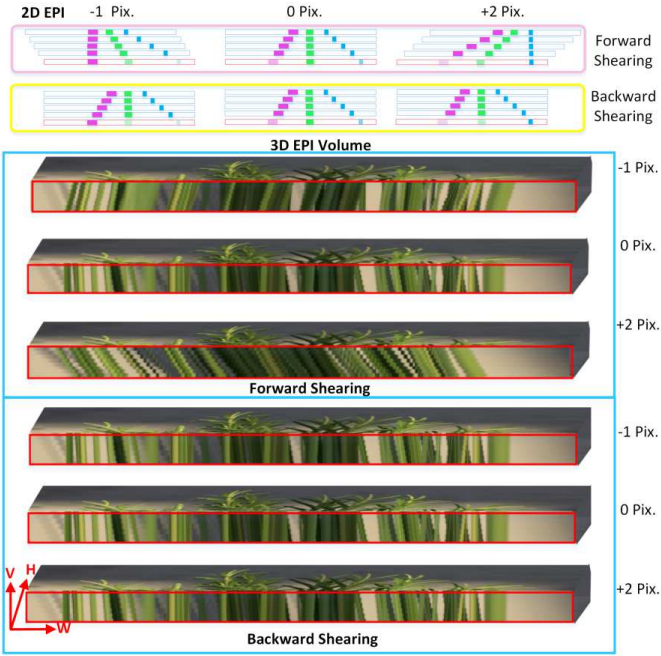}
\caption{Difference between shearing 2D EPI image and 3D EPI volume.}
\label{fig:4}
\end{figure}

\subsection{The forward and backward shearing strategy}
Since we can only have discrete angular samples with existing light field devices, discontinuous line structures in EPI pose difficulties to both view synthesis and depth estimation methods.
In order to analyze 1D structures in 2D EPI images, the authors in ~\cite{Suzuki2016} and ~\cite{Wu2019} proposed a shearing and inverse shearing strategy. In this paper, we use a similar shearing strategy that we extend to 3D EPI volumes.
The 3D EPI volumes are formed by stacking the different views in a line.
Each 3D EPI volume is then sheared using different disparity (or shearing) values. The number of sheared EPIs should be related to the disparity range. For example, if the disparity range is $[-K, +K]$, and the number of sheared EPI candidates is higher than $2K+1$, then at least one of these candidates will correspond to the right shear. Please note that shearing the 3D EPI volumes can be seen as constructing plane sweep volumes \cite{Kim2013}, but taking a volume of images rather than one single image, thus favouring consistency across views (see Figure ~\ref{fig:4} for an illustration of the 3D EPI volumes compared with 2D EPIs).

So far, most EPI line feature analysis and extraction methods operate on 2D EPI images, which can only deal with the horizontal or vertical local spatial information. In \cite{Wang2018}, Wang \etal proposed instead to apply 3D convolutions on 3D EPI volumes, which can take advantage of both horizontal and vertical spatial information. In this paper, we further improve the 3D convolutions on 3D EPI volumes by adding forward shearing and backward shearing strategy. The multiple sheared 3D EPI volumes can lead to more precise and robust computationally angular view extrapolation.

Let $S_{u} = LF(u_0,v,x,y)$ be a 3D subset of a 4D light field when $u$ is fixed, \eg the views of a horizontal line on the angular plane. We define $\chi(S_{u},d)$ the forward shearing operation, which can generate multiple sheared EPI volumes as ${S_u(v,x,y) = S_{u}(v,x+vd_i,y), d_i\in[d_{min}, d_{max}]}$, where $[d_{min}, d_{max}]$ is the disparity range between two adjacent views. The function $\hat{\chi}(S_{u})$ will denote the backward shearing operation. As shown in Figure ~\ref{fig:4}, the 4D input of the proposed network is composed of all these concatenated sheared 3D EPI volumes.

\subsection{Details on network implementation}

To apply the above forward and backward shearing strategy, the proposed EPI-SENet network is functionally divided into two parts, the EPI extrapolation network and the confidence fusion network. Both the EPI extrapolation and the confidence fusion networks are based on the Encoder-Decoder U-Net architecture. The details of the proposed EPI-SENet architecture are given in Table \ref{Tab:1}.

\begin{table}[h]
\centering
\caption{Structure of the EPI-SENet}
\label{Tab:1}
\small
\begin{tabular}{c|c|c|c|c|c}
\hline
Layers & Kernels & Strides & filters & Dimensions & Repeat \\
\hline
     \multicolumn{6}{c}{Input: 4D\_V = $\chi(S_{u},d)$, shape = [B,S,H,W,V,C]}\\
\hline
     \multicolumn{6}{c}{3D\_V = Map\_FN (4D\_V)}\\
\hline
    conv1 & $3x3x3$ & [1,1,1] & 8 & $H,W,V,8$ & 2x\\
\hline
    conv\_d1 & 3x3x3 & [2,2,2] & 16 & H/2,W/2,V/2,16 & 1x\\
\hline
    conv2 & 3x3x3 & [1,1,1] & 16 & H/2,W/2,V/2,16 & 2x\\
\hline
    conv\_d2 & 3x3x3 & [2,2,2]& 32 & H/4,W/4,V/4,32 & 1x\\
\hline
    conv3 & 3x3x3 & [1,1,1] & 32 & H/4,W/4,V/4,32 & 2x\\
\hline
    deconv1 & 3x3x3 & [2,2,2] & 16 & H/2,W/2,V/2,16 & 1x\\
\hline
    \multicolumn{6}{c}{concat1 = deconv1 + conv2}\\
\hline
    conv4 & 3x3x3 & [1,1,1] & 16 & H/2,W/2,V/2,16 & 2x\\
\hline
    deconv2 & 3x3x3 & [2,2,2] & 8 & H,W,V,8 & 1x\\
\hline
    \multicolumn{6}{c}{concat2 = deconv2 +conv1}\\
\hline
    conv5 & 3x3x3 & [1,1,1] & 8 & H,W,V,8 & 2x\\
\hline
    conv\_d3 & 3x3x3 & [1,1,2] & 16 & H,W,V/2,16 & 1x\\
\hline
    $V^{c}_{i}$ & 3x3x3 & [1,1,1] & 1 & H,W,V/2,1 & 1x\\
\hline
    \multicolumn{6}{c}{$S^\prime_{u} = 4D\_V +  V^{c}_{i} +  V^{c}_{i}$} \\
\hline
    \multicolumn{6}{c}{Input: $Ext\_4D\_V$ = $\hat{\chi}(S^\prime_{u},d)$, shape = [B,H,W,2V,S$\times$C]}\\
\hline
    conv6 & 3x3x3 & [1,1,1] & 8 & H,W,2V,8 & 2x\\
\hline
    conv\_d3 & 3x3x3 & [2,2,2] & 16 & H/2,W/2,V,16 & 1x\\
\hline
    conv7 & 3x3x3 & [1,1,1] & 16 & H/2,W/2,V,16 & 2x\\
\hline
    conv\_d4 & 3x3x3 & [2,2,2]& 32 & H/4,W/4,V/2,32 & 1x\\
\hline
    conv8 & 3x3x3 & [1,1,1] & 32 & H/4,W/4,V/2,32 & 2x\\
\hline
    deconv3 & 3x3x3 & [2,2,2] & 16 & H/2,W/2,V,32 & 1x\\
\hline
    \multicolumn{6}{c}{concat3 = deconv3 +conv7}\\
\hline
    conv4 & 3x3x3 & [1,1,1] & 16 & H/2,W/2,V,16 & 2x\\
\hline
    deconv4 & 3x3x3 & [2,2,2] & 8 & H,W,2V,8 & 1x\\
\hline
    \multicolumn{6}{c}{concat4 = deconv4 +conv6}\\
\hline
    conv9 & 3x3x3 & [1,1,1] & 16 & H,W,2V,16 & 2x\\
\hline
    conv\_d5 & 3x3x3 & [1,1,2] & 8 & H,W,V,8 & 1x\\
\hline
    conv\_d6 & 3x3x3 & [1,1,2] & 16 & H,W,V/2,16 & 1x\\
\hline
    conv10 & 3x3x3 & [1,1,1] & S & H,W,V/2,S & 1x\\
\hline
    \multicolumn{6}{c}{$\omega_i$ = Softmax(conv10,S), shape = [B,H,W,V/2,S]}\\
\hline
    \multicolumn{6}{c}{output = $\omega_i \times v^{c}_{i}$}\\
\hline
\end{tabular}
\end{table}

After forward shearing $\chi(\circ)$, if we ignore the training batch dimension $B$ and the color dimension $C$, the input data volume is 4D, the four dimensions corresponding to the shear $S$, the spatial height $H$ and width $W$ of each view, and the number of angular views $V$.
In order to learn line features present in the 3D EPI volumes of dimension $[H,W,V]$, the EPI extrapolation network applies independent and parallel 3D convolutions on the 3D EPI volumes of dimension $[H,W,V]$ for each shear.

This network aims at extrapolating the 3D EPI volumes of dimension $[H,W,V]$ for each shear, as
\begin{equation}\label{equ:equ8}
V^{c} = Extrap(\chi(S_{u},d_i)), i\in [0, S]
\end{equation}
where $i\in [0, S=2K+1]$ denotes the different shear values, and $\chi(.)$ denotes the shearing operation. The quantity $V^{c}$ represents the set of extrapolated candidates.
The 3D EPI volumes, comprising original views and extrapolated ones $S^\prime_{u}$, are then backward sheared with $\hat{\chi}(\circ)$, for each shear value $i$, thus forming a 4D volume (4D\_V), where the operation $'+'$ stands for the concatenation operation in Table \ref{Tab:1}. The fusion network then merges the extrapolated 3D EPI volumes for the different shears to produce the final extrapolated view $V$ as
\begin{equation}\label{equ:equ9}
\begin{split}
V & = Fusion(\chi(S_{u},d), V^{c}) \\
  & = \sum_{i=1}^{S} \omega_i \times v^{c}_i \qquad where \quad v^{c}_i \in V^{c}
\end{split}
\end{equation}

Those extrapolated candidates are given a high confidence weight, when their shearing corresponds to the actual depth. This fusion process can handle EPI discontinuities by combining different shearings.
The input of the fusion network is formed by concatenating the extrapolated views and the original input views, as denoted by $S^\prime_{u}$ in Table \ref{Tab:1}. Please notice that, for each shearing, two copies of extrapolated views $V^{c}$ are concatenated with the original input 4D shearing volume $4D\_V$, so that the downsampling and upsampling by a factor of 2 performed by the fusion encoder-decoder network based on U-net does not lead to unnecessary information loss. Then, all extended EPI shearings are sequentially stacked in the channel dimension, so that the 3D convolution can still be applied on the $[H,W,V]$ dimension. In the fusion stage, the $conv\_d5$ and $conv\_d6$ layers reduce the
$V$ dimension by a factor $2$, so that the fusion network generates an output containing half the number of input views. The final output is computed as a linear combination of all extrapolated shears $v^{c}_{i}$ with weights given by a confidence score $\omega_i$.

For each iteration, a new 3D volume is formed by the views extrapolated at the previous iteration stacked with half of the previous set of input views. For example, let $\{V^{\prime}_{i+5},V^{\prime}_{i+6}\}$ be the extrapolated views from the input set $\{V_{i+1},V_{i+2},V_{i+3},V_{i+4}\}$. Then, the input of the next iteration will be $\{V_{i+3},V_{i+4},V^{\prime}_{i+5},V^{\prime}_{i+6}\}$. Taking and shearing this 3D volume as a new 4D input, the extrapolation can be iteratively performed along this direction.

The proposed network can predict or extrapolate a number of views which is half the number of input views, \eg, it will output 2 novel views when taking $4$ views as input. The extrapolation can be applied in both the horizontal and vertical dimensions. To perform the vertical extrapolation, the input views should be rotated by $90$ degrees, then the vertical disparities will be treated as horizontal disparities. So, we can use the same network for the extrapolation in both directions.

\section{Experiments}
In this section,  we validate the effectiveness of the proposed the light field baseline extension framework, and conduct experiments on the refocusing precision enhancement. The quantitative and qualitative comparisons are tested on multiple public datasets of both synthetic light fields, and real light fields captured by 1st and 2nd generation Lytro cameras, as well as on light field microscopy datasets. Please note that lens distortions may yield unsatisfying errors, but in most public datasets, the views are extracted using tools (\eg the light field toolbox \cite{Decoding2013Dan}), which cope with lens distortions.

\subsection{The EPI-SENet training}
In the training stage, we first train the extrapolation network independently, then the fusion network is trained while fine tuning the extrapolation network in an end-to-end training phase. Thanks to the use of multiple shears and of confidence scores for the fusion, our network does not need accurate depth estimation.

\begin{table*}[htb]
\centering
\caption{PSNR and SSIM comparison of extrapolation results}
\label{Tab:2}
\small
\begin{center}
\scalebox{0.8}{
\begin{tabular}{| c | c | c | c | c | c | c | c | c |}
\hline
\textbf{Mean value} & \multicolumn{2}{ c |}{\textbf{FDL\cite{LePendu2019}}} & \multicolumn{2}{ c |}{\textbf{MPI\cite{Mildenhall2019}}} & \multicolumn{2}{ c |}{FPFR\cite{Shi2020Learning}} & \multicolumn{2}{ c |}{\textbf{Ours}}  \\
\cline{2-9}
\textbf{PSNR (dB)/(SSIM)} & \textbf{1.6X} & \textbf{2.3X}& \textbf{1.6X} & \textbf{2.3X}& \textbf{1.6X} & \textbf{2.3X} & \textbf{1.6X} & \textbf{2.3X}\\
\hline
HCI Syn. $[-2.56,+2.71]$ & 29.804/(0.860)  & 28.373/(0.846) & \textbf{41.023/(0.990)} & \textbf{39.230/(0.986)} & 39.163/(0.987) & 37.805/(0.983) & 40.371/(0.985) & 37.197/(0.973) \\ \hline
INRIA Syn. $[-1.76,+3.38]$ & 22.665/(0.765)  & 21.257/(0.742) & 36.956/(0.988) & 34.854/(0.982) & \textbf{37.171/(0.990)} & \textbf{35.362/(0.987)} & 36.970/(0.984) & 33.558/(0.970) \\ \hline
EPFL Illum $[-1.60,+1.70]$ & 23.410/(0.839)  & 22.674/(0.820) & 33.318/(0.981) & 30.381/(0.963) & 30.889/(0.967) & 29.416/(0.954) & \textbf{34.991/(0.983)} & \textbf{31.705/(0.967)} \\ \hline
INRIA Lytro $[-1.39,+1.64]$ & 21.595/(0.673)  & 20.299/(0.628) & 25.628/(0.781) & 23.748/(0.740) & 25.175/(0.777) & 23.375/(0.744) &\textbf{28.303/(0.901)} & \textbf{24.465/(0.781)} \\  \hline
INRIA Illum $[-2.77,+5.61]$ & 24.545/(0.779)  & 23.423/(0.749) & 27.339/(0.877) & 26.169/(0.853) & 27.300/(0.880) & 26.420/(0.866) &\textbf{31.146/(0.942)} & \textbf{27.909/(0.883)} \\  \hline
Microscopy $[-2.22,+2.60]$ & 23.196/(0.716)  & 21.378/(0.684) & 29.018/(0.886) & 25.586/(0.835) & 26.142/(0.850) & 23.953/(0.823) &\textbf{31.111/(0.903)} & \textbf{27.075/(0.851)} \\  \hline
\end{tabular}}
\end{center}
\end{table*}
The training loss function is defined as
\begin{equation}\label{equ:equ10}
\begin{split}
loss & = \frac{1}{N^{2}} \sum_{x=1, y=1}^{N} \left\|V_{gt}(x,y)-V_{pred}(x,y)\right\| \\
     & + \frac{\gamma}{N^{2}} \sum_{x=1, y=1}^{N} \left\|\nabla V_{gt}(x,y)- \nabla V_{pred}(x,y)\right\|
\end{split}
\end{equation}
where $V_{gt}$ and $V_{pred}$ are the ground truth view and the extrapolated view respectively, $\nabla$ denotes the gradient operator, and $N$ is the width and height of the predicted view.
To preserve sharp textures, in the experiments, we used $\gamma = 2.0$ to balance the mean of absolute differences and the mean of gradient differences.

We trained the EPI-SENet model on grayscale images, and the model has been used to process each color channel of color images independently. The training process takes less than 10 hours on a Tesla P100 with 16GB memory. The initial learning rate has been set to $0.0001$ for the first $200$ epochs, then it has been decreased by half every next $200$ epochs. We used the Adam optimizer\cite{Kingma2014}, and set the numerical stability parameter $\epsilon = 0.0001$. After training, we performed the extrapolation on a Nvidia 2080Ti with 11GB memory, which is a general and widely available platform.

In order to perform a quantitative comparison on real scenes, we used $four$ views on a row to extrapolate $two$ views along the same row. Since the shearing value corresponds to the disparity range, when the disparity range is limited to $[-3.0, +3.0]$, the network input volume is of dimension $[8,7,64,64,4,1]$, where the numbers represent the batch size (8), number of shearings (7), patch size (64, 64), and the number of input views respectively. We used the INRIA synthetic dense light field datasets\cite{Jiang2017} for training, including $'Ballon\_coucou'$, $'Bike\_dense'$,$'Big\_clock'$,$'Microphone'$, $'Flying\_toys'$. The proposed model is trained only on these five groups of training sets, and have been used for extrapolating all the structured light fields of the INRIA synthetic, lytro and illum datasets\cite{Jiang2017}, HCI datasets\cite{HCIdata2013}, EPFL datasets\cite{EPFLdata2016} and Stanford microscopy datasets\cite{Levoy2006LFM}, which further shows the robustness and versatility of the proposed framework.

\subsection{Extrapolation results}
\begin{figure}[htbp]
\centering
\includegraphics[width=\linewidth]{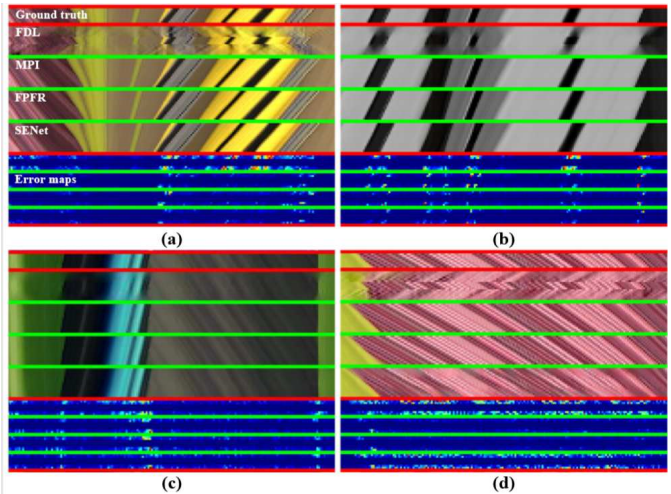}
\caption{Extrapolated EPIs obtained with different methods. (a)-(c) are EPI extrapolation results using the $Still\_life$, $Buddha$ and $Butterfly$ light fields respectively, (d) shows a case where the extrapolation is not accurate due to the presence of subtle and repetitive structures. In each sub-image, the ground truth EPI (8-rows), extrapolated EPIs (16-rows) are shown when using FDL\cite{LePendu2019}, MPI\cite{Mildenhall2019}, FPFR\cite{Shi2020Learning} and the proposed extrapolation respectively. In the bottom, we show error maps (8-rows) of the different extrapolation results (difference with the ground truth). We can see that the proposed extrapolation outperforms the state-of-the-art methods, except for the subtle and repetitive structures, \eg the case in (d).}
\label{fig:5}
\end{figure}

\begin{figure}[htbp]
\centering
\includegraphics[width=\linewidth]{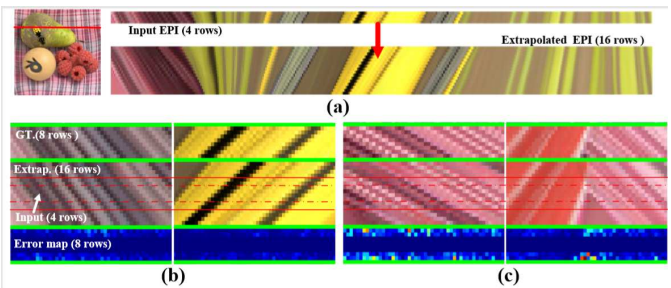}
\caption{Extended EPI results using the proposed SENet extrapolation.
(a) The input EPI and Extrapolated EPI, which correspond to the red marked line. (b) From top to bottom are $8$-rows (or views) ground truth EPIs, the extrapolated $16$-rows EPIs and error maps computed as the difference between the ground truth and the extrapolated EPIs. In the extrapolated $16$-rows EPIs, the middle 4-rows are input EPIs before extrapolation (the parts between the thin dashed red lines). Each error map includes $8$-rows,  because we only have $8\times8$ ground truth views (between the thin red lines). (c) EPI extrapolation cases showing that our extrapolation is vulnerable to subtle and repetitive structures (left sub-image), or to fully occluded regions (right sub-image).}
\label{fig:6}
\end{figure}

We first show visual examples of EPI extrapolation results in Figure ~\ref{fig:6}. We selected $8$ views in a line from the $'Still life'$ scene (HCI \cite{HCIdata2013}) as ground truth. In Figure ~\ref{fig:6}(a), the input EPI is composed of the same row from the middle $4$ views, then the $4$ rows are extrapolated to $16$ rows. The proposed algorithm performs well in most cases, and the angular consistency is well preserved, as shown in Figure ~\ref{fig:6}(b). But, this may not be the case in presence of subtle and repetitive structures or occluded regions, \eg as shown in Figure ~\ref{fig:6}(c). In Figure ~\ref{fig:6} (b) and (c), since we have only $8$-views ground truth, the error maps show the differences between $8$ rows of extrapolated views (between the two thin dashed red lines) and ground truth views. We can see that angular consistency is well preserved by the proposed algorithm. Figure \ref{fig:5} shows EPIs extrapolated with different methods.

To show the effectiveness of each module of the proposed extrapolation, we have conducted an ablation study by removing different modules, as shown in Table \ref{Tab:5}. With the fusion process, the performance of the proposed algorithm is improved, especially when applying a $4.0X$ extrapolation. When removing both the forward and backward shearing operations, the PSNR decreases by more than 6dB. When removing only the backward shearing (keeping the forward shearing), the PSNR remains acceptable, it means that the forward shearing is more important than the backward shearing.

\begin{table*}[htbp]
\centering
\caption{Ablation study of the proposed SENet pipeline}
\label{Tab:5}
\footnotesize
\begin{center}
\begin{tabular}{| c | c | c | c | c | c | c |}
\hline
\textbf{Mean value} & \multicolumn{2}{ c |}{Synthetic datasets} & \multicolumn{2}{ c |}{Lytro $\&$ Illum datasets} & \multicolumn{2}{ c |}{Mircoscopy datasets}  \\
\cline{2-7}
\textbf{PSNR (dB)/(SSIM)} & \textbf{1.6X} & \textbf{2.3X}& \textbf{1.6X} & \textbf{2.3X}& \textbf{1.6X} & \textbf{2.3X}\\
\hline
Complete pipeline & 38.825/(0.985) & 35.543/(0.972) & 30.778/(0.934) & 27.291/(0.859) & 31.111/(0.903) & 27.705/(0.851) \\ \hline
Without fusion & 38.427/(0.969) & 34.992/(0.932) & 30.233/(0.917) & 26.673/(0.831) & 30.536/(0.892) & 26.934/(0.840) \\ \hline
Without backward-shearing & 37.544/(0.951) & 33.085/(0.936) & 29.903/(0.890) & 24.687/(0.828) &	27.881/(0.846) & 24.489/(0.827) \\ \hline
Without shearing & 32.029/(0.928) & 27.954/(0.844) & 23.614/(0.822) & 20.286/(0.713) & 24.912/(0.829) & 21.062/(0.740) \\ \hline
\end{tabular}
\end{center}
\end{table*}

Then, we compare the extrapolation results of the proposed EPI-SENet extrapolation method with the ones obtained with the Fourier disparity layer (FDL) based extrapolation method of \cite{LePendu2019}, with the MPI-based methods of \cite{Mildenhall2019} and \cite{Srinivasan2019}, and with the FPFR method \cite{Shi2020Learning}.

In both \cite{Mildenhall2019} and \cite{Srinivasan2019}, the authors render novel views by forward warping and alpha compositing of MPI layers. In order to compare our proposed extrapolation method designed for structured light fields, we have adapted the camera pose estimation module of \cite{Mildenhall2019} for this case, and then finetuned the network for view extrapolation.

\begin{figure*}[htbp]
\centering
\includegraphics[width=0.9\linewidth]{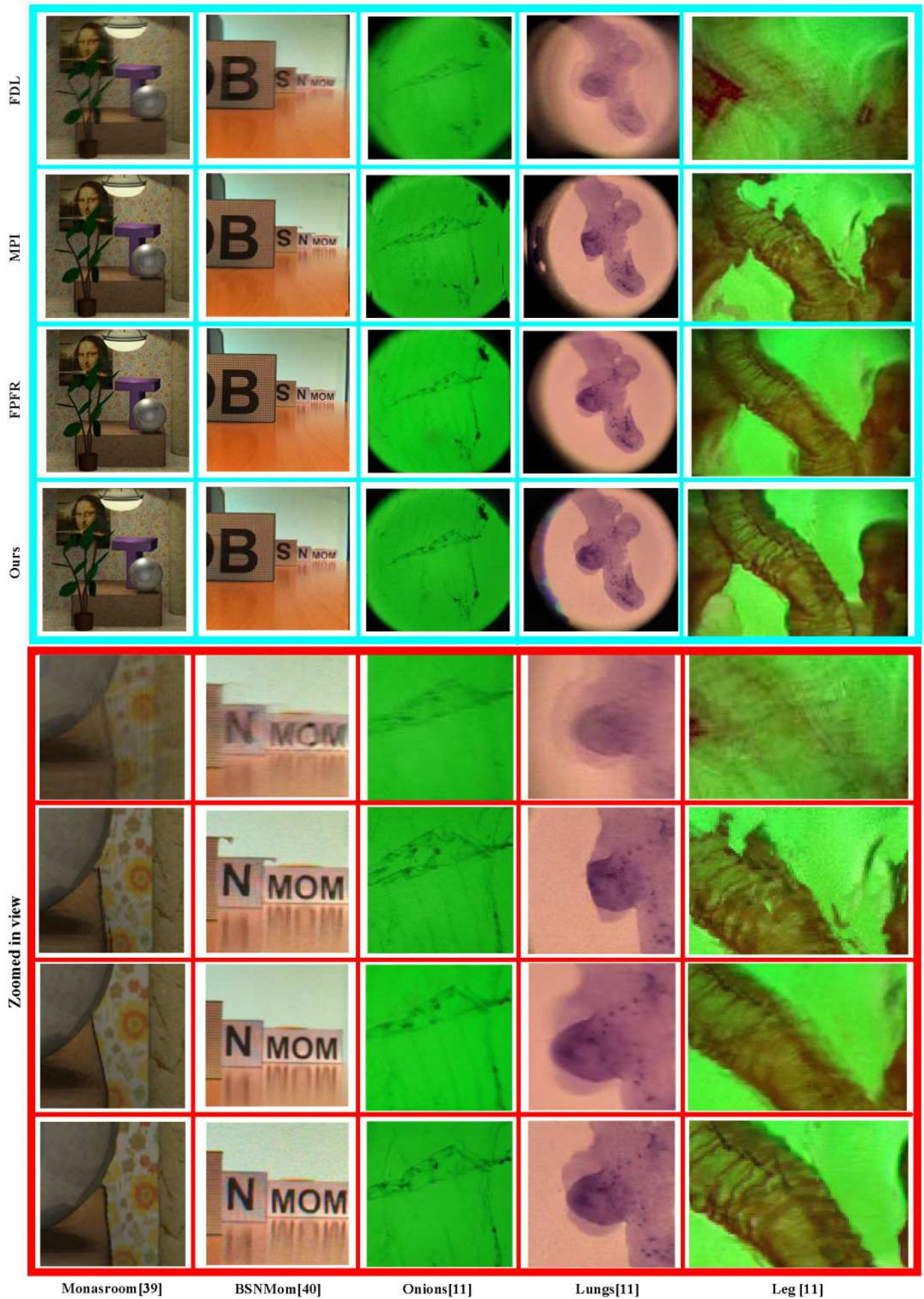}
\caption{Visual comparison of extrapolation results obtained with four different methods, FDL \cite{LePendu2019}, MPI-based extrapolation\cite{Mildenhall2019}, FPFR-based extrapolation \cite{Shi2020Learning}, and ours, with a 4X larger baseline. For five light field datasets, we show that our extrapolation method can generate much better results than the methods in \cite{LePendu2019}, \cite{Mildenhall2019} and \cite{Shi2020Learning}. From the zoomed views, we can see that the FDL method may introduce blur, and the MPI method may generate wrong textures if regions of ambiguous or repetitive texture are assigned with inaccurate opacity weights when using the multiple layer representation. The FPFR method may introduce blur when the depth estimation is not accurate enough, \eg in the scene of $Onions$ and $Leg$}
\label{fig:7}
\end{figure*}

For light fields having $8\times8$ views, we use the central $4\times4$ views to extend the baseline in both the vertical and horizontal directions. PSNR and SSIM values are computed between the extrapolated views and the two ground truth views that exist on each side of the input views considered in the test. The disparity range of each dataset is given, where the disparities of Lytro, Illum and microscopy datasets are estimated using \cite{Shi20depth}. As shown in Table \ref{Tab:2}, with synthetic datasets, the proposed extrapolation method gives results comparable to those obtained with the MPI-based extrapolation. However, due to the narrow baseline and the presence of noise, the MPI-based method gives less accurate results with real Lytro light field datasets (see Fig.~\ref{fig:7}). Please note that Figure \ref{fig:7} shows extrapolated results after three iterations of the algorithm, i.e. with a baseline $4.0\times$ larger than the original baseline. From Figure \ref{fig:7}, we can see that the FDL method introduces blur, which can result from the use of inaccurate depth when computing the layered representation. The second row of Figure \ref{fig:7} shows the MPI extrapolation results with light field microscopy datasets. The background and foreground are wrongly reconstructed for the scene $'Mouse\_lungs'$ and $'Insect\_leg'$, which can badly impact the extrapolation results. For example, the left part of mouse lungs is covered by the pink background, while the insect leg boundaries are extended. For the first two datasets of Figure \ref{fig:7}, one can observe that the FPFR extrapolation can generate better results on fine textures and object boundaries, while having lower background noise. The FPFR learning employs pixel-wise reconstruction and feature-based reconstruction to improve both low and high frequencies. Different from the proposed algorithm, both pixel-wise reconstruction and feature-based reconstruction of FPFR needs an accurate scene depth prior. If the depth estimation is not very accurate, the boundaries and texture will be also inaccurately reconstructed, \eg in the third row of Figure \ref{fig:7} (please see the zoomed views of $Onion$ and $Leg$).
In general, the MPI or FPFR-based extrapolations are more sensitive to depth estimation, while the proposed algorithm is more precise and robust when extrapolating light fields of dense and complex scenes. Indeed, in the case of complex scenes, the axial or depth resolvability given by the MPI or FPFR-based extrapolation methods may not be precise enough, especially when targeting a $2.0X$ magnification of the baseline of light field microscopy. More details of PSNR and SSIM comparison of commonly used synthetic and real scenes can be found in Table \ref{Tab:3}, where we show the PSNR and SSIM obtained for each scene independently.

\begin{table*}[tbh]
\centering
\caption{PSNR and SSIM of extrapolation results with different methods and several datasets.}
\label{Tab:3}
\small
\begin{center}
\scalebox{0.8}{
\begin{tabular}{| c | c | c | c | c | c | c | c | c |}
\hline
\textbf{Mean value} & \multicolumn{2}{ c |}{\textbf{FDL\cite{LePendu2019}}} & \multicolumn{2}{ c |}{\textbf{MPI\cite{Mildenhall2019}}} & \multicolumn{2}{ c |}{FPFR\cite{Shi2020Learning}} & \multicolumn{2}{ c |}{\textbf{Ours}} \\
\cline{2-9}
\textbf{PSNR (dB)/(SSIM)} & \textbf{1.6X} & \textbf{2.3X}& \textbf{1.6X} & \textbf{2.3X}& \textbf{1.6X} & \textbf{2.3X}& \textbf{1.6X} & \textbf{2.3X}\\
\hline
Bikes\cite{EPFLdata2016}& 21.414/(0.780) & 20.730/(0.754)& 32.286/(0.978) & 28.782/(0.955)& 30.202/(0.961) & 28.217/(0.943) & \textbf{34.444/(0.982)} & \textbf{30.586/(0.962)}\\ \hline
Friends\cite{EPFLdata2016} & 25.405/(0.899) & 24.617/(0.885)& 34.349/(0.985) & 31.979/(0.971)& 31.576/(0.972) & 30.615/(0.964) & \textbf{35.538/(0.985)} & \textbf{32.825/(0.972)}\\ \hline
Bicycle\cite{HCIdata2013}& 21.075/(0.694) & 20.648/(0.688)& 33.694/(0.975) & \textbf{31.819/(0.967)}&  32.406/(0.972)& 30.838/(0.964) &  \textbf{34.617/(0.976)} & 31.575/(0.958)\\ \hline
Boxes\cite{HCIdata2013}& 31.693/(0.943) & 29.056/(0.912)& 37.673/(0.981) & 34.751/(0.972)& 34.467/(0.975)& 32.115/(0.965) & \textbf{38.296/(0.983)} & \textbf{35.355/(0.974)}\\ \hline
Buddha\cite{HCIdata2013}& 31.033/(0.926) & 29.788/(0.908)& 45.007/(0.997) & \textbf{43.534/(0.995)}& 43.553/\textbf{(0.997)}&42.461/(0.995)& \textbf{45.184}/(0.996) & 42.499/(0.994)\\ \hline
Butterfly\cite{HCIdata2013}& 35.377/(0.963) & 34.154/(0.957)& 42.693/\textbf{(0.994)} & \textbf{41.995/(0.993)} & \textbf{42.803}/(0.993)& 41.568/(0.992) & 42.357/(0.988) & 40.453/(0.984)\\ \hline
Dino\cite{HCIdata2013}& 33.829/(0.956) & 31.588/(0.939)& 43.258/(0.995) & 41.712/(0.993)& 43.162/(0.995)& \textbf{41.688/(0.994)} & \textbf{44.226/(0.995)} & 41.443/(0.991)\\ \hline
MonasRoom\cite{HCIdata2013}& 29.486/(0.890) & 28.873/(0.877)& 44.023/(0.996) & \textbf{41.944/(0.994)} & 42.637/\textbf{(0.995)} & 41.313/(0.993) & \textbf{44.074}/(0.993) & 40.395/(0.988)\\ \hline
Stilllife\cite{HCIdata2013}& 17.360/(0.520) & 17.223/(0.504)& 34.597/(0.981) & 33.082/(0.975) & \textbf{35.833/(0.985)} & \textbf{34.649/(0.982)} & 30.871/(0.960) & 26.783/(0.912)\\ \hline
BouquetFlower2\cite{Jiang2017}& 27.079/(0.848) & 25.628/(0.819)& 28.213/(0.866) & 26.893/(0.843)& 27.759/(0.867)& 26.778/(0.853) & \textbf{31.434/(0.932)} & \textbf{27.926/(0.858)}\\ \hline
Bumblebee\cite{Jiang2017}& 14.543/(0.469) & 13.845/(0.428)& 24.830/(0.888) & 22.355/(0.836)& 25.110/(0.891) & 23.303/(0.857) & \textbf{28.721/(0.937)} & \textbf{25.263/(0.886)}\\ \hline
Field\cite{Jiang2017}& 25.798/(0.872) & 24.762/(0.837)& 26.771/(0.895) & 26.200/(0.882)& 26.822/(0.901) & 26.373/(0.892) & \textbf{32.233/(0.971)} & \textbf{29.027/(0.939)}\\ \hline
Leaves\cite{Jiang2017}& 27.259/(0.900) & 25.957/(0.877)& 28.253/(0.916) & 27.548/(0.907)& 28.282/(0.921) & 27.659/(0.916) & \textbf{32.115/(0.962)} & \textbf{29.348/(0.930)}\\ \hline
Toys\cite{Jiang2017}& 28.049/(0.806) & 26.923/(0.784)& 28.628/(0.818) & 27.848/(0.795)& 28.526/(0.822) & \textbf{27.989/(0.812)} & \textbf{31.229/(0.909)} & 27.978/(0.804)\\ \hline
Beers\cite{Jiang2017}& 22.203/(0.677) & 20.370/(0.606)& 22.973/(0.707) & 21.484/(0.651)& 22.630/(0.701) & 21.335/(0.664) & \textbf{26.020/(0.883)} & \textbf{21.771/(0.718)}\\ \hline
BSNMom\cite{Jiang2017}& 23.618/(0.739) & 22.238/(0.701)& 27.009/(0.806) & 24.749/(0.771)& 26.280/(0.803) & 24.236/(0.775) & \textbf{30.145/(0.914)} & \textbf{25.612/(0.800)}\\ \hline
Guitar\cite{Jiang2017}& 18.569/(0.637) & 17.375/(0.580)& 25.002/(0.808) & 23.187/(0.773)& 24.947/(0.811) & 23.162/(0.780) & \textbf{27.391/(0.913)} & \textbf{24.396/(0.814)}\\ \hline
TapeMeasure\cite{Jiang2017}& 21.990/(0.640) & 21.211/(0.625)& 27.528/(0.801) & 25.574/(0.765)& 26.842/(0.794) & 24.767/(0.758) & \textbf{29.657/(0.894)} & \textbf{26.083/(0.792)}\\ \hline
Dinosaur\cite{Jiang2017}& 17.910/(0.659) & 17.326/(0.647)& 32.353/(0.972) & 29.390/(0.951) & \textbf{34.501/(0.984)} & \textbf{32.169/(0.978)} & 29.219/(0.944) & 25.931/(0.895)\\ \hline
Flowers\_clock\cite{Jiang2017}& 19.793/(0.766) & 19.189/(0.759)& \textbf{41.060/(0.994)} & \textbf{39.159/(0.992)} & 40.388/(0.994) & 39.128/(0.992) & 39.060/(0.992) & 35.917/(0.986)\\ \hline
Kitchen\_board\cite{Jiang2017}& 27.563/(0.923) & 24.965/(0.871)& 35.125/(0.988) & 34.093/(0.987)& 35.282/(0.988) & 34.068/(0.986) & \textbf{37.684/(0.994)} & \textbf{34.969/(0.991)}\\ \hline
Smiling\_crowd\cite{Jiang2017}& 15.759/(0.513) & 15.080/(0.489)& 35.017/(0.992) & 32.820/(0.988)& 35.297/(0.991) & 32.840/(0.986) & \textbf{37.327/(0.994)} & \textbf{33.013/(0.988)}\\ \hline
White\_roses\cite{Jiang2017}& 32.298/(0.962) & 29.727/(0.944)& 41.224/(0.994) & \textbf{38.807}/(0.992) & 40.403/(0.994) & 38.604/\textbf{(0.993)} & \textbf{41.560/(0.994)}& 37.960/(0.989)\\ \hline
Fluorcrayons\cite{Levoy2006LFM}& 22.949/(0.717) & 22.284/(0.699)& 31.927/\textbf{(0.927)} & 29.196/\textbf{(0.899)}& 29.203/(0.901) & 27.706/(0.886) & \textbf{33.503}/(0.890) & \textbf{30.786}/(0.853) \\ \hline
Interleaved\cite{Levoy2006LFM}& 18.912/(0.693) & 18.161/(0.655)& 27.125/\textbf{(0.951)} & 22.606/(0.881)& 25.574/(0.925) & 23.266/\textbf{(0.888)} & \textbf{29.456}/(0.924) & \textbf{24.515}/(0.879) \\ \hline
Leg\cite{Levoy2006LFM}& 20.783/(0.644) & 18.271/(0.588)& 24.326/(0.769) & 19.901/(0.692)& 23.523/(0.757) & 20.659/(0.720) & \textbf{27.318/(0.847)} & \textbf{21.921/(0.760)}\\ \hline
Meanbug\cite{Levoy2006LFM}& 27.377/(0.848) & 24.000/(0.783)& 31.252/(0.927) & 27.301/\textbf{(0.890)}& 27.605/(0.915) & 24.749/(0.886) & \textbf{32.323/(0.916)} & \textbf{28.458}/(0.884)\\ \hline
Mosaiced\cite{Levoy2006LFM}& 18.814/(0.416) & 18.155/(0.426)& 25.393/(0.834) & 21.539/(0.701)& 23.751/(0.808) & 20.803/(0.723) & \textbf{28.943/(0.903)} & \textbf{23.345/(0.784)}\\ \hline
Mouselungs\cite{Levoy2006LFM}& 28.107/(0.858) & 24.927/(0.828)& 32.154/(0.917) & 29.699/(0.912) & 29.225/\textbf{(0.921)} & 26.792/\textbf{(0.918)} & \textbf{33.052}/(0.918) & \textbf{29.739}/(0.907)\\ \hline
Onion\cite{Levoy2006LFM}& 25.433/(0.837) & 23.854/(0.808)& 30.948/(0.875) & 28.858/(0.866)& 24.115/(0.716) & 23.698/(0.743) & \textbf{33.184/(0.921)} & \textbf{30.764/(0.894)}\\ \hline
Total Mean & 23.879/(0.746) & 22.523/(0.720) & 31.642/(0.888) &	 29.503/(0.865) & 30.386/(0.879) & 28.824/(0.864) & \textbf{33.032}/\textbf{(0.917)} & \textbf{29.629}/\textbf{(0.874)}\\ \hline
\end{tabular}}
\end{center}
\end{table*}

\subsection{Noise and iterative extrapolation analysis}

We have tested the proposed extrapolation method in presence of different levels of salt-and-pepper noise and Gaussian noise.
The results are given in Table \ref{Tab:8} and Table \ref{Tab:9}. In Table  \ref{Tab:8}, $NP$ represents the percentage of noisy pixels among the whole set of pixels, and $\sigma$ stands for the standard variation of the Gaussian distribution, which is zero mean in the experiments. Since the proposed extrapolation method exploits angular consistency in EPIs, it performs well in presence of weak noise. Line structures in EPIs are deteriorated when increasing the noise, which leads a performance decrease.

Table \ref{Tab:6} shows how the estimation error accumulates along the iterations when performing iterative extrapolation, using the $17\times17$ stanford $Bunny$ and $Lego\_Knights$ light fields. When proceeding iteratively, line structures can be deteriorated when previously predicted pixels are not accurate, hence estimation errors can accumulate. However, the error remains acceptable when the number of iterations is small. For this reason, we suggest applying the extrapolation up to a 4.0X baseline extension factor.

\begin{table*}[tbh]
\centering
\caption{PSNR/SSIM performance in presence of salt-and-pepper noise.}
\label{Tab:8}
\small
\begin{center}
\scalebox{0.85}{
\begin{tabular}{| c | c | c | c | c | c | c | c | c |}
\hline
\textbf{NP(\%)} & Bicycle & Boxes & Buddha & Butterfly & Dino & MonasRoom & Stilllife & Mean value\\
\hline
0.1 & 32.565/(0.962) & 34.589/(0.964) & 35.939/(0.977) & 35.148/(0.966) & 37.266/(0.978) & 35.521/(0.972) & 29.293/(0.948) & 34.332/(0.967) \\ \hline
0.5 & 28.989/(0.913) & 29.488/(0.899) & 29.956/(0.915) & 29.125/(0.886) & 30.543/(0.907) & 29.485/(0.902) & 27.107/(0.905) & 29.242/(0.904) \\ \hline
1.0 & 26.650/(0.856) & 27.004/(0.836) &	27.277/(0.849) & 26.425/(0.804) & 27.811/(0.835) & 26.778/(0.828) & 25.450/(0.858) & 26.771/(0.838) \\ \hline
1.5 & 25.262/(0.808) & 25.510/(0.779) &	25.590/(0.791) & 24.903/(0.740)	& 26.259/(0.774) & 25.085/(0.763) & 24.136/(0.814) & 25.249/(0.781) \\  \hline
2.0 & 24.226/(0.764) & 24.260/(0.723) &	24.432/(0.739) & 23.747/(0.682) & 25.164/(0.725) & 24.182/(0.719) & 23.220/(0.772) & 24.176/(0.732) \\  \hline
\end{tabular}}
\end{center}
\end{table*}

\begin{table*}[tbh]
\centering
\caption{PSNR/SSIM performance in presence of Gaussian noise.}
\label{Tab:9}
\small
\begin{center}
\scalebox{0.85}{
\begin{tabular}{| c | c | c | c | c | c | c | c | c |}
\hline
\textbf{$\sigma$} & Bicycle & Boxes & Buddha & Butterfly & Dino & MonasRoom & Stilllife & Mean value\\
\hline
5 & 31.765/(0.944) & 34.011/(0.947) & 34.559/(0.956) & 35.284/(0.946) &	35.577/(0.957) & 34.669/(0.948) & 28.700/(0.931) & 33.509/(0.947) \\ \hline
10 & 28.424/(0.876) & 29.889/(0.866) & 29.922/(0.877) &	30.786/(0.864) & 30.330/(0.865) & 30.093/(0.867) & 26.377/(0.869) & 29.403/(0.869) \\ \hline
20 & 24.182/(0.723) & 25.246/(0.689) & 25.336/(0.712) &	26.070/(0.693) & 25.154/(0.665) & 25.431/(0.698) & 22.802/(0.721) & 24.889/(0.700) \\ \hline
30 & 21.592/(0.594) & 22.549/(0.552) & 22.628/(0.573) &	23.220/(0.556) & 22.156/(0.507) & 22.700/(0.561) & 20.541/(0.594) & 22.198/(0.562) \\  \hline
\end{tabular}}
\end{center}
\end{table*}

\begin{table}[tbh]
\centering
\caption{PSNR/SSIM performance along the iterations.}
\label{Tab:6}
\small
\begin{center}
\scalebox{0.8}{
\begin{tabular}{| c | c | c | c | c | c | c |}
\hline
\textbf{Iterations} & \textbf{1st} & \textbf{2nd} & \textbf{3rd} & \textbf{4th} & \textbf{5th} & \textbf{6th}\\
\hline
\textbf{PSNR(dB)} & 41.211  & 39.657 & 35.215 & 29.432 & 24.112 & 17.241 \\ \hline
\textbf{SSIM} & 0.993  & 0.985 & 0.981 & 0.896 & 0.769 & 0.634 \\ \hline
\end{tabular}}
\end{center}
\end{table}

\subsection{Refocusing precision evaluation}
After the extrapolation, the digital refocusing can produce a shallower DoF due to the extended baseline, thus leading to more accurate refocusing. On the microscopy datasets, by using the central $4\times4$ views as our inputs and the central $8\times8$ as ground truth, we compared the PSNR between refocusing results with ground truth refocusing, as shown in Table \ref{Tab:4}. Our algorithm obviously outperforms state-of-the-art methods in \cite{Mildenhall2019} and \cite{Shi2020Learning}, with most datasets and a $4.0X$ baseline extension, except for the $'Mouselungs'$ scene, for which the FPFR and MPI based PSNR results are slightly higher. Instead of comparing on a single focal plane, Table \ref{Tab:4} shows the mean PSNR value of $61$ refocus planes within the disparity range $[-3.0, +3.0]$.

\begin{table}[tbh]
\centering
\caption{PSNR comparison of refocusing results in the disparity range $[-3.0, +3.0]$ with a 4.0X baseline.}
\label{Tab:4}
\small
\begin{center}
\begin{tabular}{| c | c | c | c | c |}
\hline
\textbf{Mean PSNR (dB)} &\textbf{FDL\cite{LePendu2019}} & \textbf{MPI\cite{Mildenhall2019}} & \textbf{FPFR\cite{Shi2020Learning}} & \textbf{Ours}  \\
\hline
Fluorcrayons &32.077  & 40.250 & 40.208 &\textbf{41.254}  \\ \hline
Interleaved & 29.425  & 34.981 & 35.311 &\textbf{36.825}  \\ \hline
Leg & 28.538  & 30.550 & 31.023 &\textbf{32.644} \\ \hline
Meanbug & 37.739  & 39.992 & 39.997 &\textbf{41.459} \\  \hline
Mosaiced &  28.563  & 33.362 & 33.214 &\textbf{34.895} \\  \hline
Mouselungs & 39.367 & 42.153 & \textbf{42.307} &41.849 \\  \hline
Onion & 39.498 & 41.739 & 39.796 &\textbf{42.625} \\  \hline
\end{tabular}
\end{center}
\end{table}

\begin{figure*}[htbp]
\centering
\includegraphics[width=\linewidth]{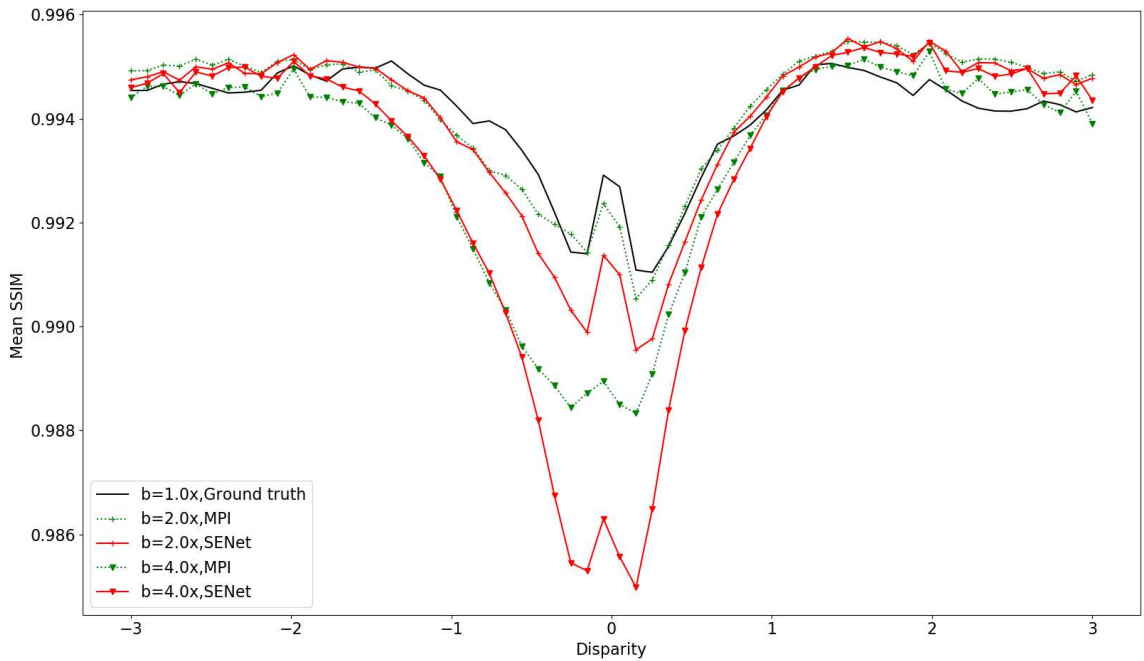}
\caption{SSIM between two adjacent refocus images with 1.0X, 2.0X, 4.0X baseline extension factors for two methods, the MPI-based method \cite{Mildenhall2019} and the proposed one. Lower is the SSIM, more distinguishable (i.e. showing more structural differences) are the two adjacent refocus planes.}
\label{fig:8}
\end{figure*}

\begin{figure*}[htb]
\centering
\includegraphics[width=\linewidth]{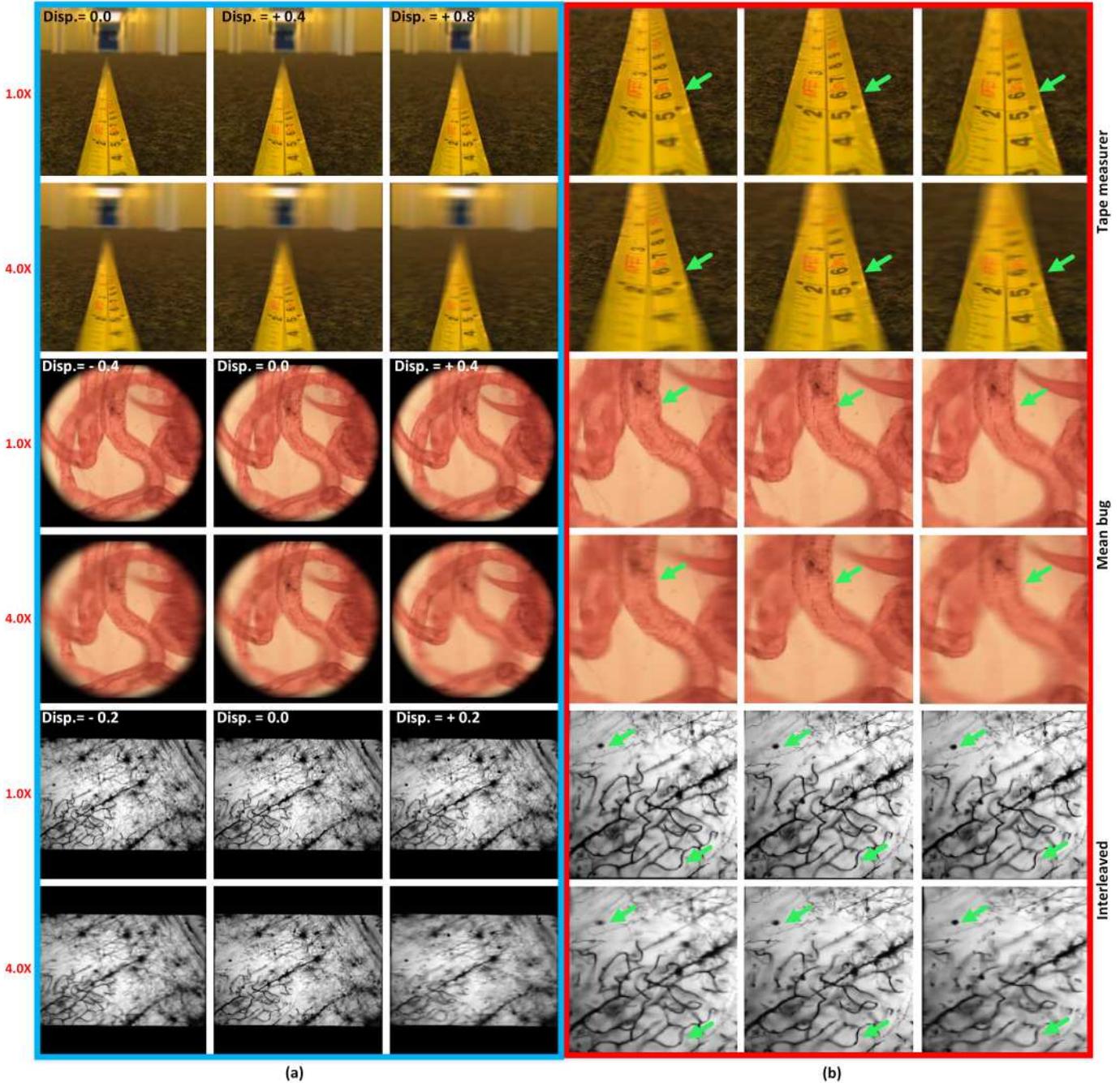}
\caption{Comparison of axial refocusing results with 1.0X, 4.0X baselines, where different disparity values correspond to different focal planes in axial direction. (a) Refocused results on same axial focus planes, with $1.0X$ and $4.0X$ baselines respectively. (b) Zoomed views, the green arrows are pointing at details on different refocusing planes that are more distinguishable with a $4.0X$ baseline and less resolvable with the $1.0X$ baseline.}
\label{fig:9}
\end{figure*}

To validate the axial refocusing enhancement, we further tested the resolvability or distinguishability of the adjacent refocusing results. The SSIM is employed to measure the differences between two adjacent refocus images. In Figure \ref{fig:8},  for each axial focal plane (corresponding to different disparities), we give the mean SSIM of $7$ microscopy datasets. A small value of mean SSIM indicates high distinguishability. The black curve stands for distinguishability for each axial focal plane with the ground truth $1.0X$ baseline light field, while the other curves marked with cross and triangles represent the axial distinguishability of refocused results with $2.0X$ and $4.0X$ baselines respectively. In Figure \ref{fig:8}, the curves of MPI and ours are shown in green color and red color respectively, from which we can see that, the proposed axial precision enhancement can give better performances than the MPI-based approach. The vertical peak in Figure \ref{fig:8} can be explained by the non-linear characteristics of the disparity space. A same disparity unit stands for a larger physical distance between the distinghuishable adjacent refocus planes, when moving away from the zero disparity plane. A larger physical separation will lead to high distinguishability (smaller SSIM value), which yields the central peak of SSIM curves. The left and right wings of the curves are due to the fact that the entire specimen is out of focus, the blurriness indeed leads to a high SSIM value. Finally, Figure \ref{fig:9} visually compares refocused images with $1.0X$ and $4.0X$ baseline. The detail changes can hardly be recognized with $1.0X$ baseline original refocusing, \eg, those areas pointed by the green arrow. But, one can easily distinguish the sharp focused plane of these details on the refocused images with $4.0X$ baseline using the proposed framework.

\subsection{Extrapolation complexity}

In Table \ref{Tab:7}, we compare the computational complexity of three learning based extrapolation methods in terms of FLOPs, number of parameters and runtimes. Although the MPI algorithm has the lowest FLOP values, the proposed algorithm takes only 1.1 seconds to predict each view. The MPI algorithm \cite{Mildenhall2019} takes 4.7 seconds, and the FPFR (pixel-feature fused reconstruction) algorithm \cite{Shi2020Learning} takes 5.4 seconds for the same prediction task (including the time of data augmentation by rotation and flipping). Including the pre-and-post processing, our algorithm is the best in terms of run-time performance, because the proposed algorithm can generate multiple output views (half number of input views) in each prediction step. In addition, the operations carried out for shearing and extrapolation are highly parallelizable. The proposed network is also lighter in terms of number of parameters (290,848 parameters) than the MPI network (681,581 parameters), and the FPFR network (14,313,505 parameters).

\begin{table}[tbh]
\centering
\caption{Extrapolation complexity comparison of three learning-based methods.}
\label{Tab:7}
\small
\begin{center}
\begin{tabular}{| c | c | c | c |}
\hline
Complexity & \textbf{MPI\cite{Mildenhall2019}} & \textbf{FPFR\cite{Shi2020Learning}} & \textbf{Ours}  \\
\hline
Calculation(Mflops) & 1.39 & 33.26 & 12.17 \\ \hline
Parameters(M) & 0.68 & 14.31 & 0.29 \\ \hline
Run-time(s) & 4.7 & 5.4 & 1.1 \\ \hline
\end{tabular}
\end{center}
\end{table}

\subsection{Limitations}
\begin{figure}[htbp]
\centering
\includegraphics[width=\linewidth]{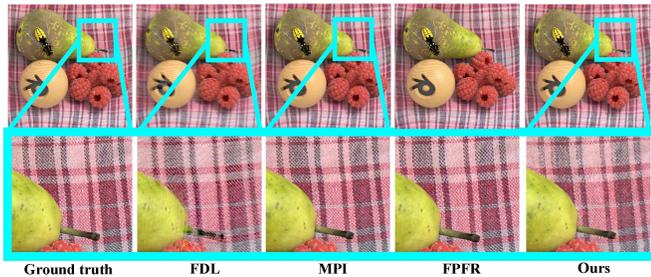}
\caption{Extrapolation results for a scene having subtle and repetitive structures.}
\label{fig:10}
\end{figure}

Since the method relies on EPI line structures, our input should be structured views, which is a limitation of the proposed framework. The proposed extrapolation method is vulnerable in the case of occluded regions, of regions with non-lambertian reflection, or with subtle and repetitive structures. For example, this can be observed in the background tablecloth in Figure \ref{fig:10} (see the blur in the last column). Another limitation is that our extrapolation is less accurate when applying a large number of iterative extrapolations. For this reason, we suggest applying the extrapolation within a $4.0X$ baseline extension, which is enough for specific applications, \eg in light field microscopy.

\section{Conclusion}
In this paper, we introduced a learning-based extrapolation for enhancing the axial resolution, when performing digital re-focusing. The method allows us to extend the baseline to $4.0X$ larger for structured light fields. It can handle EPI discontinuous structures by applying a forward and backward shearing strategy on 3D EPI volumes. When compared with existing methods, the view extrapolation can generate better results, when applied to uniformly sampled and structured light fields. After the baseline extension, on light field microscopy datasets, the refocusing precision can be significantly improved, which validates the effectiveness of the proposed framework. The proposed axial precision enhancement is suitable for those applications that require accurate refocusing precision. Due to the theoretical differences, the proposed approach does not need explicit or accurate depth estimation. The proposed method is more suitable for structured light fields with small baselines as those captured by plenoptic cameras (especially for the plenoptic 1.0 cameras). It is better if the number of shearings can cover the entire disparity range. Note also that non uniform disparity distribution as in the case of unstructured light fields, as well as lens distortion, image abberations, may lead to unsatisfying errors. The definition of a quantitative metric to measure axial resolution could be the scope of further study. One could also investigate how to introduce constraints for further enforcing angular consistency.

\section*{Acknowledgment} {This work has been funded in part by the National Natural Science Foundation of China (No. 61871319 and No. 62031023), in part by the Natural Science Basic Research Plan of Shaanxi Province(No.2019JM-221), in part by the China Scholarship Council (CSC) under grant No.201808610055, and in part by the EU H2020 Research and Innovation Programme under grant agreement No 694122 (ERC advanced grant CLIM)}

\bibliographystyle{IEEEtran}
\bibliography{axial_sr_bibliography}

\end{document}